# Spatiotemporal programming of shape memory polymers for tunable vibration responses

Samantha Averitt[1], Jay Sim[1], Ruike Renee Zhao[1]*

[1]Department of Mechanical Engineering, Stanford University, Stanford, CA 94305, USA
*Corresponding author. Email: rrzhao@stanford.edu

**Abstract:** Stimuli-responsive materials offer a promising route to dynamically tailor material behavior, enabling adaptive functionalities in changing environments. However, simultaneously implementing spatial and temporal control in active material systems remains a fundamental challenge. Here, we present a photothermally programmable shape memory polymer (SMP) metamaterial platform that enables dynamic control of elastic wave propagation through reconfigurable spatiotemporal temperature distributions. By projecting user-defined light patterns onto carbon-doped SMP (C-SMP) surfaces, spatial thermomechanical property landscapes are generated, enabling a wide range of elastic wave functionalities, including tunable transmission spectra and programmable vibrational modes. By treating both space and time as design parameters, we access an expanded design space that allows a bulk C-SMP surface to function as a highly adaptive metamaterial platform. This work establishes spatiotemporal property modulation as a powerful framework for adaptive vibration control, tunable phononic systems, and other advanced dynamic applications.

## Introduction

Material systems with dynamically tunable properties have emerged as a powerful platform for creating adaptive structures that can respond to changing operational environments. These active systems enable on-demand modulation of functionality, which is critical for applications ranging from soft robotics[1,2] and adaptive aerospace structures[3,4] to tunable acoustic devices[5] and vibration mitigation[6]. Such adaptability is typically achieved through stimuli-responsive materials, which alter their physical properties in response to external triggers. For instance, materials have been engineered to modulate mechanical stiffness[7,8], optical transmission[9], and material geometry[10–13] in response to stimuli such as temperature[7], light[9,11], and electromagnetic fields[8,12,13]. Examples include shape memory polymers for mechanical reconfiguration[7], thermochromic $VO_2$ coatings with temperature-dependent changes in transparency[9], magnetorheological fluid suspensions with tunable viscosity[8], and magnetically responsive composites for untethered shape change and locomotion[12,13]. Collectively, these systems demonstrate the growing capability to tailor material behavior in real time.

Many existing metamaterial programming strategies rely on homogeneous property tuning, where the entire material undergoes uniform changes[14–16]. While effective, this paradigm fundamentally limits functionality by constraining the material to a single global state at any given time. In contrast, spatiotemporal programming of material properties provides a pathway to substantially expand the functional design space. Spatial modulation enables properties to vary throughout the material, creating heterogeneous property distributions that can be tailored to achieve functionalities such as tunable energy absorption[17], programmable mechanical responses[18], and active wave manipulation[19–21]. Incorporating temporal modulation introduces an additional degree of freedom, allowing properties to evolve in time to respond to changing environments and dynamically switch between functional states[22–26]. Combined spatiotemporal control therefore enables the reconfiguration of complex property landscapes over time, supporting highly adaptive and multifunctional metamaterial behavior. However, existing demonstrations that integrate both spatial and temporal control remain relatively limited, with many systems restricted to modulation between two discrete states[24] or to property distributions with limited spatial resolution[22,25]. This fundamentally limits the achievable number of material configurations and, consequently, the range of material properties and functionalities that can be accessed.

In this work, we leverage spatiotemporal control to create an adaptive metamaterial system with a dynamically reprogrammable vibrational response, where 'spatiotemporal' is used to describe the reprogrammability of material properties in both space and time. By continuously controlling the thermomechanical properties of carbon-doped shape memory polymer (C-SMP) surfaces across locations over time, we access an expanded design space that enables a bulk material to demonstrate a wide range of vibrational behaviors, adapting to changing environments without requiring structural redesign. As illustrated in **Fig. 1a**, user-defined light fields projected onto the material surface generate spatially and temporally varying thermal patterns. The C-SMP consists of an amorphous polymer matrix with embedded carbon nanotubes (CNTs), which provide enhanced optical absorption and enable efficient photothermal conversion. Upon illumination, absorbed light is converted into heat, creating a spatially varying temperature field that mirrors the projected light pattern. As shown in **Fig. 1b**, the elastic and viscous response of the C-SMP is dictated by temperature, quantified by the storage modulus ($E'$) and tangent delta ($\tan\delta$), respectively. Over the investigated temperature range, $E'$ spans several orders of magnitude, accompanied by a pronounced temperature-dependent variation in $\tan\delta$. Consequently, the light-induced temperature fields establish a prescribed distribution of thermomechanical properties across the material, dictating the effective mechanical response and enabling dynamic, spatially resolved control of material behavior (**Fig. 1c**). The use of grayscale light intensities enables continuous spatiotemporal tuning of the C-SMP's thermomechanical properties beyond binary 'heated' and 'unheated' states. Furthermore, programming a bulk material surface allows site-specific control independent of a predefined unit-cell architecture. To the authors' knowledge, few metamaterial systems achieve this level of continuous control over both the magnitude and spatial distribution of material properties, expanding the range of accessible configurations for adaptable vibration control. Moreover, the programmed temperature distributions are fully reversible. In the absence of light, the C-SMP cools to room temperature and returns to its initial stiffness, allowing a different property distribution to be reprogrammed to meet changing design requirements. Leveraging this highly adaptive and reprogrammable metamaterial framework, we demonstrate temporal modulation of elastic wave behavior, including tunable transmission and reprogrammable vibrational modes.

The remainder of the paper is structured as follows. The thermomechanical behavior of C-SMP is characterized to establish the foundation for thermal actuation. The relationship between these properties and the resulting elastic wave behavior is then demonstrated using a basic cantilever beam system. Subsequently, we show control of two-dimensional vibrational modes, where projector-based photothermal programming dynamically reconfigures the response of a C-SMP plate. Overall, our results demonstrate that spatiotemporal stiffness modulation in SMPs expands the programmable design space, enabling a bulk SMP surface to function as a highly reconfigurable metamaterial platform. This provides a pathway towards realizing adaptive systems capable of responding to dynamically changing environments, opening opportunities in applications such as adaptive vibration mitigation, programmable acoustic devices, and tunable phononic systems.

## Results

### Photo-thermomechanical programming of a C-SMP beam.

We first utilize a uniformly heated cantilever beam to demonstrate the programmable elastic wave transmission under varying light intensities. **Fig. 2a** illustrates the experimental setup where an illuminated C-SMP beam (5 mm × 60 mm × 1.5 mm) is subjected to out-of-plane excitation at one end. Accelerometers placed at the two ends of the sample record the acceleration signals at the input and output measurement points, denoted as $a_{in}$ and $a_{out}$, which are used to calculate the frequency-dependent transmission (see Supplementary **Note 3** and Supplementary **Fig. 5** for further details on the experimental setup and data processing). Moreover, to ensure that changes in elastic wave transmission arise from thermally induced stiffness modulation rather than deformation of the C-SMP, the beam is mounted on a 1 mm thick rigid acrylonitrile styrene acrylate (ASA) backing. **Fig. 2b** demonstrates the photothermal response of the C-SMP beam, where higher light intensities correspond to higher temperatures. When illuminated at 0, 370, 840, and 1400 W/m$^2$, the beam reaches steady-state temperatures of 22, 30, 40, and 50 °C, respectively. By varying the light intensity at the sample surface, temperature can be continuously tuned, providing a high-resolution control of the material response.

The glass transition ($T_g$) of the C-SMP defines the temperature range over which the most pronounced change in mechanical properties occurs. To ensure strong modulation between 22 °C (room temperature) and 50 °C, the C-SMP is formulated to have a $T_g$ of 26 °C[12]. As shown in **Fig. 2c**, placing the $T_g$ within this range enables substantial changes in stiffness and damping during heating, characterized by the storage modulus (*E'*) and the loss tangent (tan$\delta$), respectively. Between 22 °C and 50 °C, *E*' decreases from 70.3 to 0.7 MPa, while tan$\delta$ varies from 0.7 to 2.6. Because C-SMP is viscoelastic, its thermomechanical properties depend not only on temperature but also on loading frequency. To accurately describe the beam's dynamic response over the 10 to 70 Hz range considered here, the frequency dependence of *E*' and tan$\delta$ is also incorporated into the material model. **Fig. 2d** presents *E*' and tan$\delta$ as functions of frequency, where *E*' increases with loading frequency and tan$\delta$ reaches a maximum at ~0.4 Hz. These curves are obtained using time-temperature superposition (TTS), which shifts experimental data obtained at different temperatures to extend the effective frequency range of material characterization. The resulting TTS master curves are then fit using a generalized Maxwell model to simulate the beam's dynamic response (see Supplementary **Note 2** and Supplementary **Fig. 2** for details on the viscoelastic material characterization).

The tunable thermomechanical properties of C-SMP present a pathway to dynamically control the propagation of elastic waves. **Fig. 2e** shows the experimental transmission spectra of the C-SMP beam at steady-state temperatures of 22, 30, 40, and 50 °C, spanning the material's transition across its $T_g$. At room temperature (22 °C), the resonant frequency of the beam is 35.5 Hz, identified by the peak of the transmission curve. As the beam is heated to 30, 40, and 50 °C, its resonant frequency shifts downward to 28.8, 14.6, and 13.6 Hz, demonstrating that photothermal heating reduces the beam's stiffness and consequently lowers its resonant frequency. Moreover, temperature simultaneously modulates the resonant amplitude through changes in material damping. As the temperature rises from 22 to 30, 40, and 50 °C, the amplitude at resonance shifts from 10.2 to 5.7, 3.6, and 6.7 dB. This non-monotonic evolution is directly tied to the temperature-dependent damping of the C-SMP, where the tan$\delta$ increases as the material approaches its glass transition and subsequently decreases at higher temperatures. Note that the observed vibrational response is reversible and shows good repeatability over multiple heating and cooling cycles (Supplementary **Figs. 7** and **8**). **Fig. 2f** presents the corresponding simulated transmission spectra for the C-SMP beam. These simulations reproduce the same qualitative trends measured experimentally, including a downward shift in resonance with increasing temperature and non-monotonic changes in damping. For instance, the simulations show an increase in dissipation from 22 °C to 40 °C, followed by a decrease in dissipation from 40 °C to 50 °C, consistent with the experimentally observed evolution of tan$\delta$. Although the numerical model shows some discrepancies with experiment (see Supplementary **Note 4** for further discussion), the simulations still capture the dominant physics of the system, validating that the thermomechanical properties of C-SMP can be used to program the vibrational response. Collectively, these results demonstrate a deterministic mapping between temperature, thermomechanical properties, and elastic wave transmission in C-SMPs.

**Tuning of 2D vibrational modes.**

Photothermal actuation is now extended to a two-dimensional system in the form of a C-SMP plate, using homogeneous heating to control the vibrational modes. **Fig. 3a** shows the experimental setup, where a 100 mm × 100 mm × 1 mm C-SMP plate ($T_g$ = 26 °C) is mounted to an electrodynamic modal shaker via a rigid ASA backing (1 mm thick) and subjected to out-of-plane excitation at the location $a_{in}$. To resolve the resulting two-dimensional wave field, the acceleration is measured at multiple output points, denoted as $a_1$, $a_2$, and $a_3$, providing spatial transmission measurements that quantify elastic wave propagation across the plate. **Fig. 3b-d** present how the frequencies of the plate's vibrational modes change with temperature. As photothermal heating decreases the C-SMP's stiffness, the resonant frequency at which each vibrational mode occurs shifts downward.

In the unheated configuration (22 °C), the plate's first resonant mode occurs at 123 Hz and subsequently decreases to 84 Hz and 69 Hz upon heating to 33°C and 38 °C (**Fig. 3b**). Similarly, the frequency of the second mode shifts from 273 Hz at 22 °C to 187 Hz at 33°C and 149 Hz at 38 °C (**Fig. 3c**), while the third mode decreases from 552 Hz to 410 and 320 Hz over the same temperature range (**Fig 3d**). The accompanying simulations agree with the experimental results, reproducing the same trend of monotonically decreasing resonance with increasing temperature.

The resonant modes are identified using the experimental transmission spectra and displacement fields presented in **Fig. 3e-g**. By associating each resonant frequency with its characteristic spatial response, the evolution of a given vibrational mode can be tracked during photothermal actuation. For example, **Fig. 3e** presents the experimental transmission spectra for a plate at 22 °C, where the first resonant mode is identified by the peak at 123 Hz. At this frequency, the acceleration at the far right of the sample ($a_2$) is highest, with a transmission of 16.8 dB, followed by 10.1 dB for $a_3$, and 3.8 dB for $a_1$, indicating that the unheated plate experiences the greatest displacement near its edge and the smallest displacement at its center. Upon heating to 33 °C (**Fig. 3f**) and 38 °C (**Fig. 3g**), the peak associated with this mode decreases to 84 Hz and 69 Hz, respectively, while maintaining similar relative amplitudes across measurement points. This shift is accompanied by a reduction in transmission magnitude, showing enhanced viscoelastic dissipation in the C-SMP with increasing temperature.

This thermally programmed decrease in frequency is further validated through visualization of the displacement field. Salt particles are dispersed across the plate's surface and accumulate along the nodal lines (regions of zero displacement) during excitation, forming Chladni patterns[27,28]. Because the driving amplitude of the electrodynamic shaker is adjusted to maintain consistent visualization across frequencies (Supplementary **Note 3**), temperature-dependent changes in damping are not evident in the resulting Chladni patterns. Nevertheless, the imposed perturbations remain sufficiently small to approximate the C-SMP response as linear viscoelastic, allowing direct comparison of modal responses across temperatures. At 22 °C, the first mode consists of a curved nodal line spanning the side of the plate opposite the 123 Hz excitation. As the temperature increases, the same nodal structure is recovered at 84 Hz and 69 Hz, consistent with the decreasing frequency observed in the transmission spectra. The second and third modes exhibit analogous temperature-dependent frequency shifts while preserving their characteristic spatial responses. This tunability enables a single structure to adapt to changes in excitation frequency, maintaining a desired vibrational response without structural redesign despite variations in the driving frequency.

**Fig. 3h-j** show the simulated transmission spectra and corresponding vibrational modes of the C-SMP plate at 22, 33, and 38 °C. The simulations capture the dominant features of the temperature-dependent modal response, including the downward shift in resonance and the general structure of the displacement fields. Beyond resonant behavior, photothermal programming also enables elastic wave modulation off-resonance, providing access to intermediate standing wave patterns with hybrid modal characteristics. Supplementary **Movie 1** illustrates this behavior through the continuous evolution of Chladni patterns on a homogeneous C-SMP plate during cooling at a fixed excitation frequency. Together, these results demonstrate that thermally induced resonance shifting can accommodate changes in excitation frequency, transforming a bulk surface into a reconfigurable platform for vibration control across changing operating conditions.

**Spatial thermal programming of 2D vibrational modes.**

Patterned illumination of the C-SMP programs temperature distributions that produce spatially heterogeneous stiffness landscapes, thereby expanding the range of achievable elastic wave functionalities. **Fig. 4a** shows the vibrational response of a homogeneous C-SMP plate (100 mm × 100 mm × 1 mm) subjected to a harmonic excitation of 123 Hz at the center of an edge (red dot). Under these conditions, the plate exhibits a symmetric vibration response about its horizontal centerline. Homogeneous irradiation causes the C-SMP to undergo a uniform stiffness change, shifting the system off-resonance and producing a new standing wave pattern that preserves modal symmetry. In contrast, partial illumination creates spatially heterogeneous thermomechanical property distributions that partially reflect and scatter the incident waves. This reshapes the nodal geometries (regions of zero or low vibration amplitude), thereby expanding the range of achievable vibration responses. Four spatial thermal patterns are demonstrated in **Fig. 4b**: left half (LH), top triangle (TT), bottom triangle (BT), and middle triangle (MT). The LH and MT patterns redistribute the plate's high- and low-vibration amplitude regions while preserving symmetry about the horizontal centerline. Conversely, the TT and BT thermal patterns produce asymmetric vibration responses, enabling selective vibration localization in the bottom-left and top-left corners of the plate, respectively. The corresponding finite element analyses capture the overall features of the experimentally observed vibration responses. Notably, for the 40 °C, LH, and MT patterns, the numerical model tends to underestimate the effective stiffness of the plate, resulting in simulated displacement fields that occur at lower frequencies than those observed experimentally (Supplementary **Note 4**). Despite these frequency offsets, the simulations preserve the overall thermomechanical property distributions of the spatially programmed C-SMP, providing qualitative agreement with the experimentally observed nodal geometries. The corresponding experimental and simulated transmission spectra for the thermal patterns depicted in **Fig. 4** are provided in Supplementary **Fig. 14**, showing similar spectral features and overall trends in transmission modulation.

**Fig. 4c-d** show the displacement field of a C-SMP plate subjected to a harmonic excitation of 150 Hz at its center (red dot). For the homogeneous plate, the vibration response exhibits fourfold rotational and reflection symmetry about its center. Uniform heating preserves these symmetries while increasing the diameter of the circular nodal line relative to the unheated reference configuration (**Fig. 4c**). In comparison, spatial illumination of the bottom half (BH), two quadrants (2Q), a single quadrant (1Q), and two triangles (2T) reshapes the nodal line and introduces directional variation in the harmonic displacement field **(Fig. 4d)**. By providing directional control over the nodal geometry, spatial programming enables regions of low- and high-amplitude vibrations to be selectively repositioned, expanding the system's functional range. The resulting spatial vibration configurations are captured by the simulated displacement fields, which reproduce the experimentally observed nodal geometries. The reprogrammability of the C-SMP is further illustrated in Supplementary **Movies 2** and **3**, which show the Chladni patterns of a cantilevered and center-excited C-SMP plate, respectively, cooling from various spatially programmed thermal distributions at a fixed excitation frequency. These results demonstrate the continuous tunability of SMP, where gradual changes in stiffness and damping enable access to intermediate vibrational responses. The ability to reconfigure a system's nodal lines at a given frequency enables metamaterial adaptability in applications where the spatial distribution of vibrations must be adjusted to accommodate changing operating conditions or control objectives. By breaking modal symmetry, spatial programming expands the range of achievable vibration responses, providing a means to target specific regions for vibration suppression and localization without modifying the system's underlying structure.

**Spatiotemporal programming of 2D vibrational modes**

To modulate elastic wave functionality in response to changing environments, the transient behavior between thermal patterns must be considered alongside the steady-state response. **Fig. 5a** illustrates three distinct illumination sequences programmed onto a center-excited C-SMP plate (100 mm × 100 mm × 1 mm) over 8 min. All three sequences begin from the same initial state established under vertical illumination (Pattern 1) and converge to the same final state under horizontal illumination (Pattern 3). However, the intermediate illumination (Pattern 2), applied from $t$ = 0 to 4 minutes, differs between sequences: the light is turned off for sequence 1 (green curve), a cross pattern is projected for sequence 2 (red curve), and a horizontal pattern consistent with Pattern 3 is projected for sequence 3 (blue curve). From $t$ = 4 to 8 minutes, the same light pattern is projected for all three sequences, causing evolution toward a common material state. These transitions occur on the order of minutes, reflecting a timescale of thermal and vibrational reconfiguration larger than that of the mechanical oscillations. **Fig. 5b** presents the local temporal transmission response at 150 Hz for each sequence, measured at the point $a_{out}$ (red circle). Despite converging to comparable initial (-7.8 ± 0.2 dB, mean ± SD across sequences) and final (-13.3 ± 2.0 dB) values, the transient vibrational behavior strongly depends on the programmed thermal sequence. Sequence 1 increases the transmission at the measurement point, reaching a maximum value of -1.3 dB compared to the initial -7.7 dB at $t$ = 0. Conversely, sequences 2 and 3 result in local elastic wave attenuation, reaching minimum transmittances of -38.1 dB and -29.5 dB, respectively. The timing of these responses also varies across sequences. Sequence 1 reaches its maximum transmission at $t$ = 4 min, whereas sequences 2 and 3 reach their minimum values at $t$ = 2 min and $t$ = 1 min, respectively. Thus, time serves as an additional design parameter, enabling spatiotemporal programming to control both the nature of the elastic wave response and the time at which it occurs. Note that slight variations in the vibrational response between sequences at 0 and 8 min are expected, given the thermal-history dependence of SMP stiffness[29] (**Supplementary Note 4**).

The C-SMP's temporal response is further captured using salt particle visualization. **Fig. 5c** shows the infrared (IR) images and Chladni patterns for each thermal sequence from 0 to 8 minutes, demonstrating a clear correspondence between the evolving nodal structures and the measured transmission. At $t$ = 0 min, all three sequences share the same thermal distribution and therefore exhibit the same vibrational response, characterized by two curved vertical nodal lines. As different intermediate light patterns are applied from 0 to 4 minutes, the resulting Chladni patterns for each programming sequence progressively diverge. For instance, at $t$ = 1 min, although sequences 2 and 3 share the same cross-shaped thermal geometry, they produce distinct standing wave patterns due to spatial temperature differences: sequence 2 generates a vertically elongated elliptical nodal line, whereas sequence 3 produces a more symmetric circular nodal structure. By 4 min, each thermal sequence exhibits a distinct vibrational response. Sequence 1 produces a circular nodal line of smaller diameter, where the increased displacement near $a_{out}$ suppresses particle accumulation, consistent with the measured increase in transmission and demonstrating programmable energy localization. In contrast, the larger diameter of the nodal line in sequence 2 indicates reduced displacement at the plate's edge, illustrating tunable vibration isolation consistent with the lower measured transmittance at $a_{out}$. Sequence 3 has already approached its final thermal state by this time, reflected in the minimal evolution of Chladni patterns and the corresponding plateau in local transmission. Importantly, these transient vibrational fields correspond to different functional objectives. For instance, the energy localization observed in sequence 1 could be leveraged for applications requiring energy harvesting at the point $a_{out}$. Under different operating conditions, the vibration isolation resulting from sequence 2 could be exploited to protect sensitive components. Meanwhile, sequence 3 enables a more rapid transition between programmed vibrational states, which could be advantageous when faster adaptation is required. As the same final light pattern is applied from 4 to 8 minutes, the Chladni patterns for each sequence gradually reconverge to the same nodal structure. At $t = 8$ min, all three sequences exhibit the same horizontal temperature distribution and produce a common vibrational response characterized by four localized nodal points. Collectively, these results demonstrate temporal control over the system's vibrational modes. Engineering these transient elastic-wave fields through spatiotemporal programming has the potential to enable advanced dynamic functionalities, including programmable particle manipulation, adaptive vibration mitigation, and wave steering. The full temporal evolution of the thermal and vibrational fields is shown in Supplementary **Movie 4**, and the corresponding simulated transmission values and displacement fields are presented in Supplementary **Fig. 15**. Overall, these results demonstrate that spatiotemporal programming further expands metamaterial functionality by introducing transient regimes that cannot be captured through steady-state analysis alone. Therefore, treating both space and time as programmable dimensions provides a powerful framework for engineering time-dependent functional trajectories and achieving robust metamaterial reconfiguration in dynamically changing environments.

## Discussion

In summary, by projecting spatially and temporally varying thermal patterns onto C-SMP surfaces, we have demonstrated highly programmable control over their vibrational responses for largely enhanced adaptability. This approach allows stiffness and damping to be continuously tuned via the coupled temperature- and frequency-dependent material response, thereby realizing diverse elastic wave functionalities, including tunable transmission spectra and programmable vibrational modes. Consequently, the dynamic response can be tuned on demand by redesigning the applied thermal field without modifying the underlying material structure, enabling a bulk material to function as a highly reprogrammable metamaterial platform.

Beyond bulk material systems, extending spatiotemporal programming to SMPs with varied geometries and boundary conditions would further enrich an already vast design space, building on the established use of SMPs in geometrically reconfigurable structures[12,30–32]. Although this work focuses on elastic wave control, the presented framework can be extended to other stiffness-dependent metamaterial functionalities, including tuning the effective Poisson's ratio and stress-strain response. Moreover, the ability to program material properties in both space and time results in a high-dimensional design space that is impractical to navigate with conventional trial-and-error approaches. The thermal patterns presented here represent only a small subset of the vibration responses accessible through spatiotemporal programming. Realizing target-driven control requires efficient computational frameworks that identify the actuation parameters needed to achieve a desired vibration response. Such frameworks could facilitate the design of more complex, application-specific functionalities. For instance, resonance shifting could be leveraged to design energy harvesters that maintain a favorable vibration response over a broader frequency spectrum, thereby expanding their effective operating range. Alternatively, nodal line redistribution would be beneficial in aerospace structures that experience changing aerodynamic loads; reconfiguring the vibration response could help adapt structural dynamics across flight regimes, enhancing vibration and flutter suppression. With continued advances in inverse design and application-specific validation, the fundamental vibration capabilities of this work provide a foundation for translating spatiotemporal programming to practical systems with adaptive functionalities. Overall, we envision that spatiotemporal programming will greatly expand the functional capabilities of active metamaterials, inspiring the development of highly adaptive systems for vibration control, waveguiding, and other advanced dynamic applications.

## Methods

### Material Fabrication

The neat SMP resin contains aliphatic urethane diacrylate (Ebecryl 8807, Allnex, Alpharetta, GA), 2-phenoxyethanol acrylate (Ebecryl 114, Allnex), isobornyl acrylate (Sigma-Aldrich, St. Louis, MO), and isodecyl acrylate (Sigma-Aldrich) mixed at a weight ratio of 0.7:30.1:55:14.2. A photoinitiator (Phenylbis(2,4,6-trimethylbenzoyl)phosphine oxide, 1.5 wt%) and fumed silica (average particle size ~ 0.2 μm, Sigma-Aldrich, 2 wt%) are incorporated into the resin to enable UV-curing and to promote uniform dispersion of the carbon nanotubes. The composite SMP is prepared by adding 0.1 wt% of multi-walled carbon nanotubes (Sigma-Aldrich) to the resin matrix, followed by centrifuging for five minutes. The mixture is cast into Dragon Skin 20 (Smooth-On Inc., Macungie, PA) molds and sandwiched between two glass slides. Curing is performed under UV light for 0.5 h, followed by post-curing at 80 °C for 12 h. Additional fabrication details are provided in the Supplementary Materials.

### Mechanical Property Characterization

Uniaxial tension tests and dynamic thermomechanical measurements are conducted on a dynamic mechanical analyzer (DMA 850, TA Instruments, New Castle, DE) using 20 mm × 4 mm × 1 mm specimens. Temperature-dependent viscoelastic properties are measured by applying a 0.1% oscillatory strain while ramping the temperature from 10 to 80 °C for each $T_g$ formulation. The peak of the tan$\delta$ is used to define the glass transition temperature. Uniaxial tension tests are conducted at temperatures spanning 20 to 85 °C at a strain rate of 0.01 $s^{-1}$. The resulting stress-strain data are fit to a linear elastic model below $T_g$ and to the Arruda-Boyce model above $T_g$ to determine the temperature-dependent elastic modulus of the C-SMP. Frequency-dependent viscoelastic behavior is characterized by applying a 0.1% oscillatory strain at frequencies ranging from 0.1 Hz to 40 Hz, measuring the thermomechanical properties as functions of frequency at temperatures spanning 15 to 85 °C. The resulting storage and loss moduli are used to fit a 16-branch generalized Maxwell model. More details on material characterization are provided in the Supplementary Materials.

**Experimental Setup**

Photothermal heating is achieved using a laser projector (PowerLite L790U, Epson, Los Alamitos, CA), and the resulting temperature distributions are captured using an infrared camera (340, FOTRIC, Santa Clara, CA). Here, we define irradiance as the light intensity measured at the sample surface using a solar power meter (SM206, OEM instrument). Samples are mounted to an electrodynamic shaker (K2007E01, The Modal Shop, Cincinnati, OH) using a rigid ASA backing and excited in the out-of-plane direction using harmonic signals supplied by a waveform generator (33120A, Hewlett Packard, Palo Alto, CA). Vibrational displacements are measured using accelerometers (352A21, PCB Piezotronics, Depew, NY) placed at the point of excitation and at one or more output locations, and the resulting voltage signals are recorded using an oscilloscope (EDUX1002A, Keysight, Santa Rosa, CA). The resulting two-dimensional displacement fields are subsequently visualized by distributing salt particles (average particle size ~ 200 μm) across the sample surface. Additional experimental details are provided in the Supplementary Materials.

**Numerical Simulations**

Finite element frequency domain simulations are performed using the Solid Mechanics module in COMSOL Multiphysics 6.1. The ASA backings are modeled as a linear elastic solid with a Young's modulus of 2 GPa, mass density of 1100 kg $m^{-3}$, Poisson's ratio of 0.33, and an isotropic structural loss factor of 0.025. The C-SMP is modeled using a 16-branch generalized Maxwell model coupled with a Williams-Landel-Ferry shift function. Harmonic excitation is applied via a prescribed displacement, while all remaining boundaries are traction-free. Displacement magnitude contours are used to visualize the resulting displacement fields. Additional modeling details are provided in the Supplementary Materials.

## Data availability

All data supporting the findings of this study are available within the paper and its Supplementary Materials or from the corresponding author upon request.

## Acknowledgements

We thank N. Mourrain for assistance with experimental documentation and assembly

## Funding sources

The authors disclose support for the research of this work from the Defense Advanced Research Projects Agency (DARPA) YFA (grant number D25AC00387).

## Author contributions

R.R.Z. conceived the research and supervised the study. S.A. performed the experiments, data analysis, and FEA simulations. All authors wrote and edited the manuscript.

## Competing interests

The authors declare no competing interests.

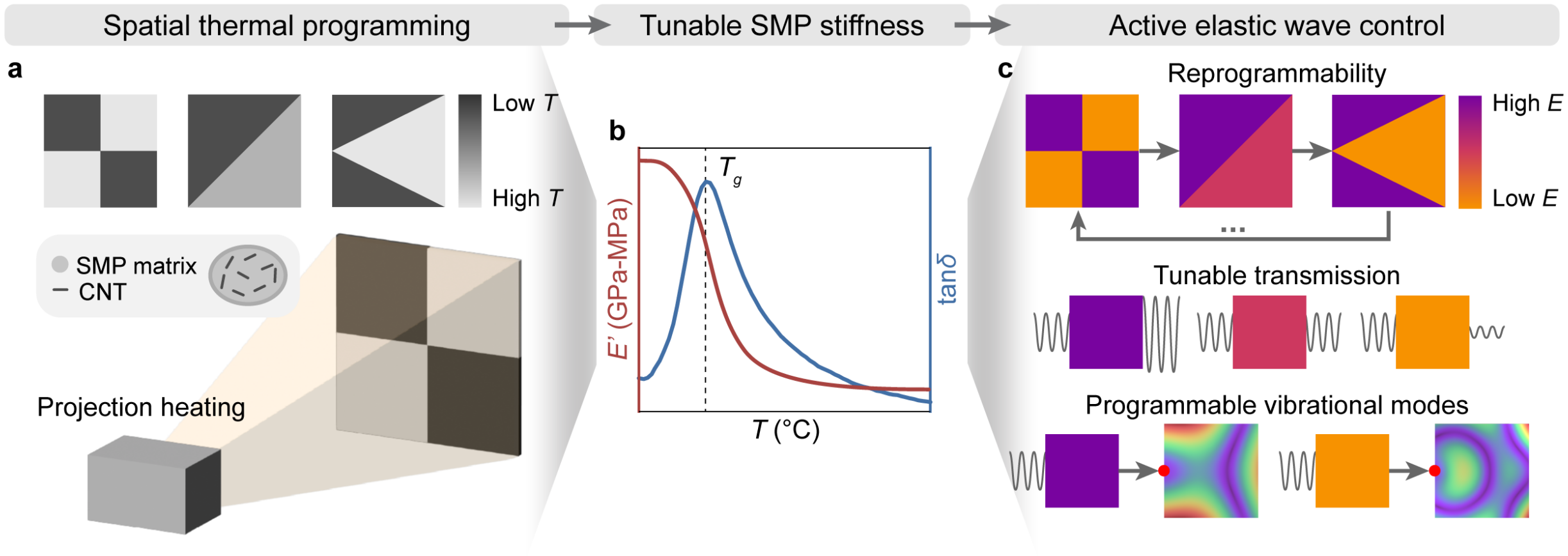

**Fig. 1: Photothermal programming of SMPs for active elastic wave control. a** User-defined thermal patterns for projector-based heating. Grayscale inputs define varying light intensities that produce spatial temperature distributions through photothermal conversion. Low light intensity corresponds to low temperature (black), and high intensity corresponds to high temperature (light gray). CNTs embedded in the polymer matrix convert incident projector light into heat. **b** Temperature-dependent stiffness of C-SMP, enabling spatial control of $E$' and tan$\delta$ through selective illumination. **c** Active elastic wave control via reprogrammable thermomechanical property distributions. Spatial stiffness control modulates the vibrational response, enabling tunable elastic wave transmission and reprogramming of vibrational modes.

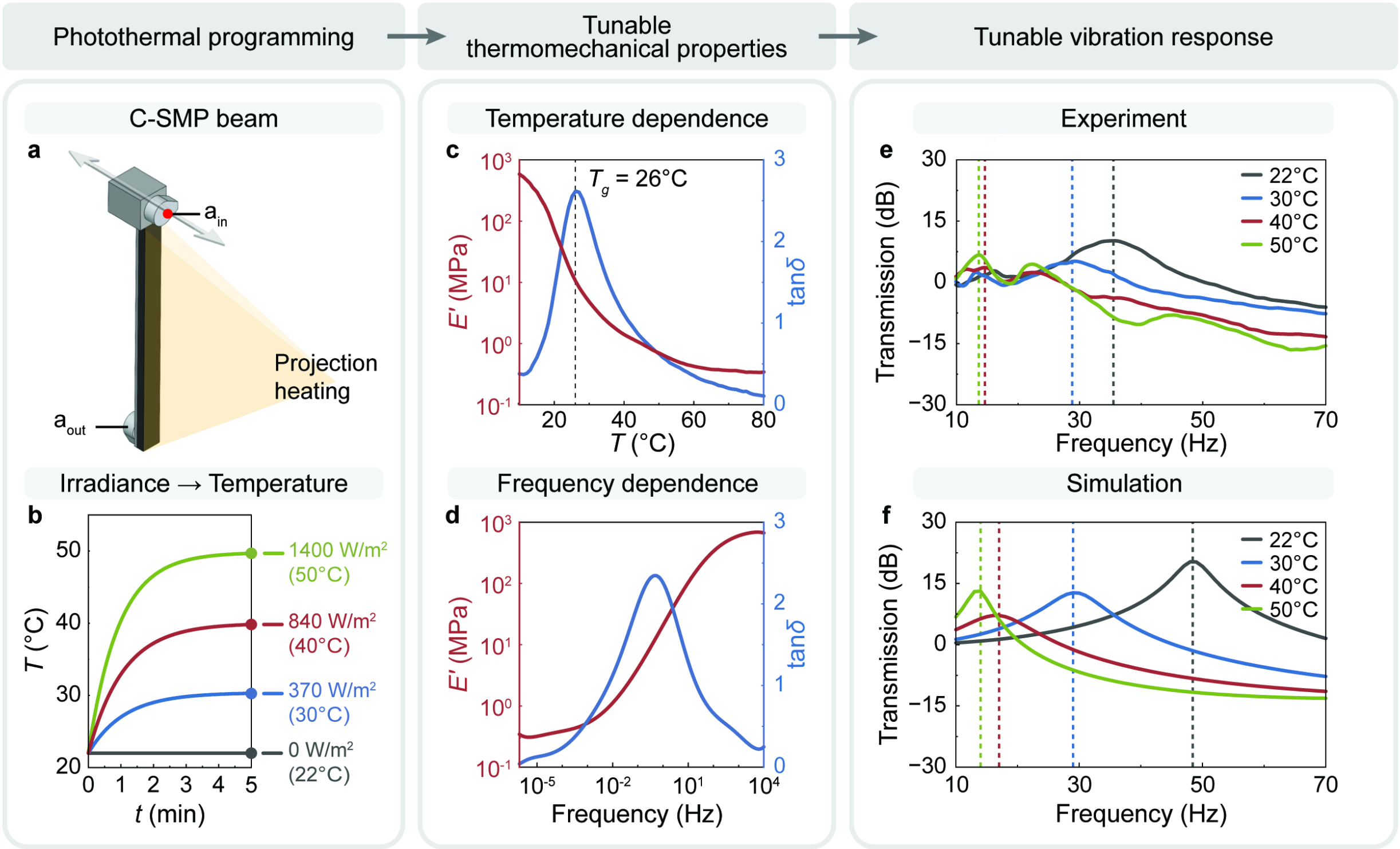

**Fig. 2: Photo-thermomechanical programming of a C-SMP beam. a** A 5 mm × 60 mm × 1.5 mm C-SMP beam is excited out-of-plane, and accelerations at input ($a_{in}$) and output ($a_{out}$) measurement points are used to determine the transmission. **b** Temperature evolution of C-SMP under photothermal irradiation at 0, 370, 840, 1400 W/m$^2$, corresponding to equilibrium temperatures of 22, 30, 40, 50 °C, respectively. **c** Temperature-dependent mechanical response of C-SMP, showing *E'* (red) and tan$\delta$ (blue) for C-SMP with a glass transition temperature of $T_g$ = 26 °C. **d** Frequency-dependent mechanical response of C-SMP, showing *E'* (red) and tan$\delta$ (blue) as functions of frequency. **e** Experimental transmission spectra evaluated at the equilibrium temperatures shown in panel b, demonstrating shifts in resonant frequency and transmission amplitude with changes in temperature. **f** Simulated transmission spectra, evaluated at the same equilibrium temperatures as in panel b.

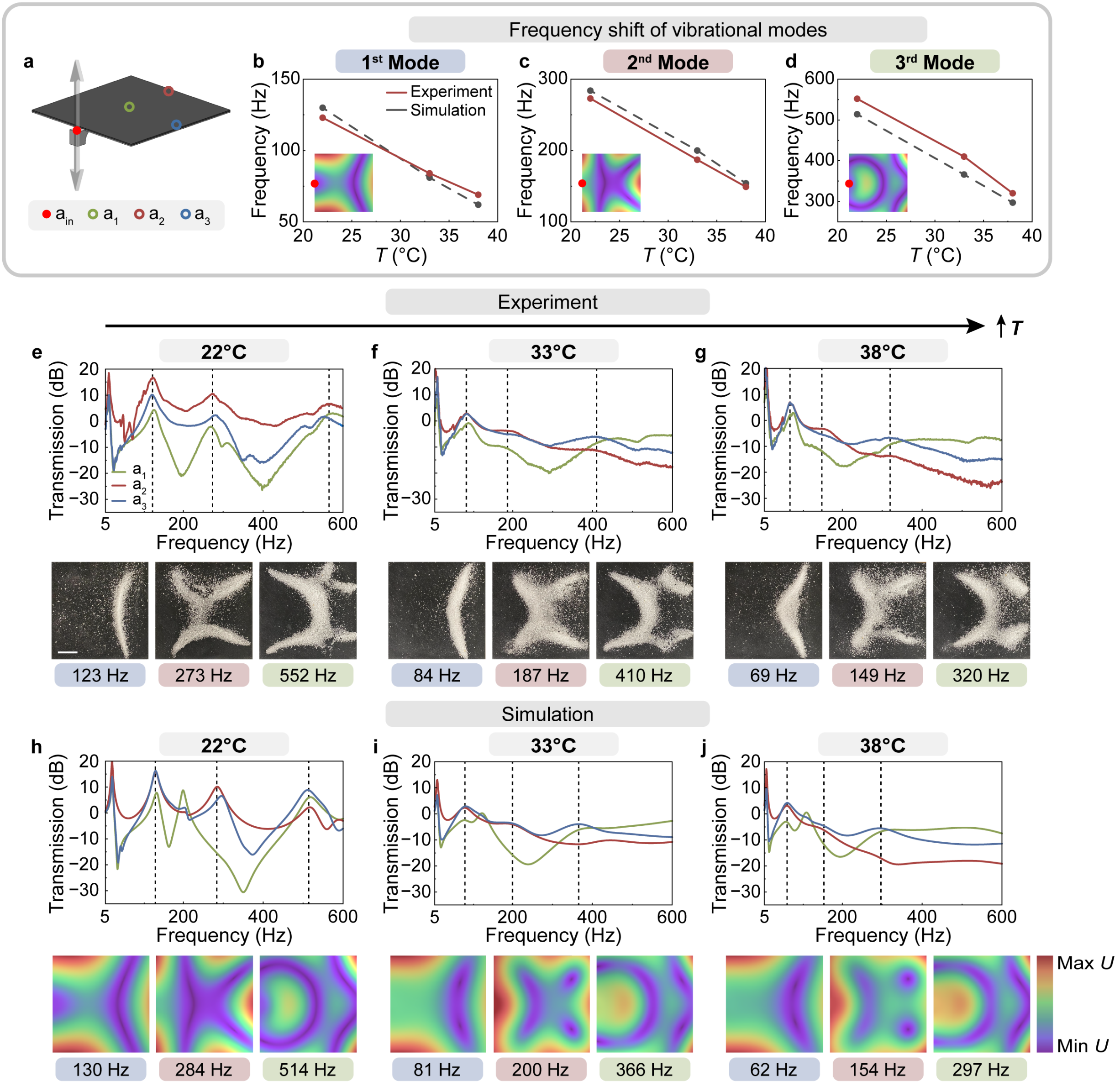

**Fig. 3: Tuning of 2D vibrational modes. a** A 100 mm × 100 mm × 1 mm C-SMP plate is excited out-of-plane, and accelerations at the input ($a_{in}$) and three output locations ($a_1$, $a_2$, and $a_3$) are measured. **b-d** Experimental (red) and simulated (black) frequencies of the (**b**) first, (**c**) second, and (**d**) third vibrational modes as functions of temperature. All modes decrease in frequency with increasing temperature due to thermally induced softening. **e-g** Experimental transmission spectra and corresponding Chladni patterns at (**e**) 22 °C, (**f**) 33 °C, and (**g**) 38 °C. Transmission is calculated from the accelerations measured at the points $a_1$ (green), $a_2$ (red), and $a_3$ (blue), providing spatial information about the resulting wave field. Resonant modes are identified from peaks in the transmission spectra (dashed lines). **h-j** Simulated transmission spectra and corresponding displacement fields at (**h**) 22 °C, (**i**) 33 °C, and (**j**) 38 °C. Purple regions indicate small displacements ($U$), whereas red regions correspond to large displacements. Scale bar: 20 mm.

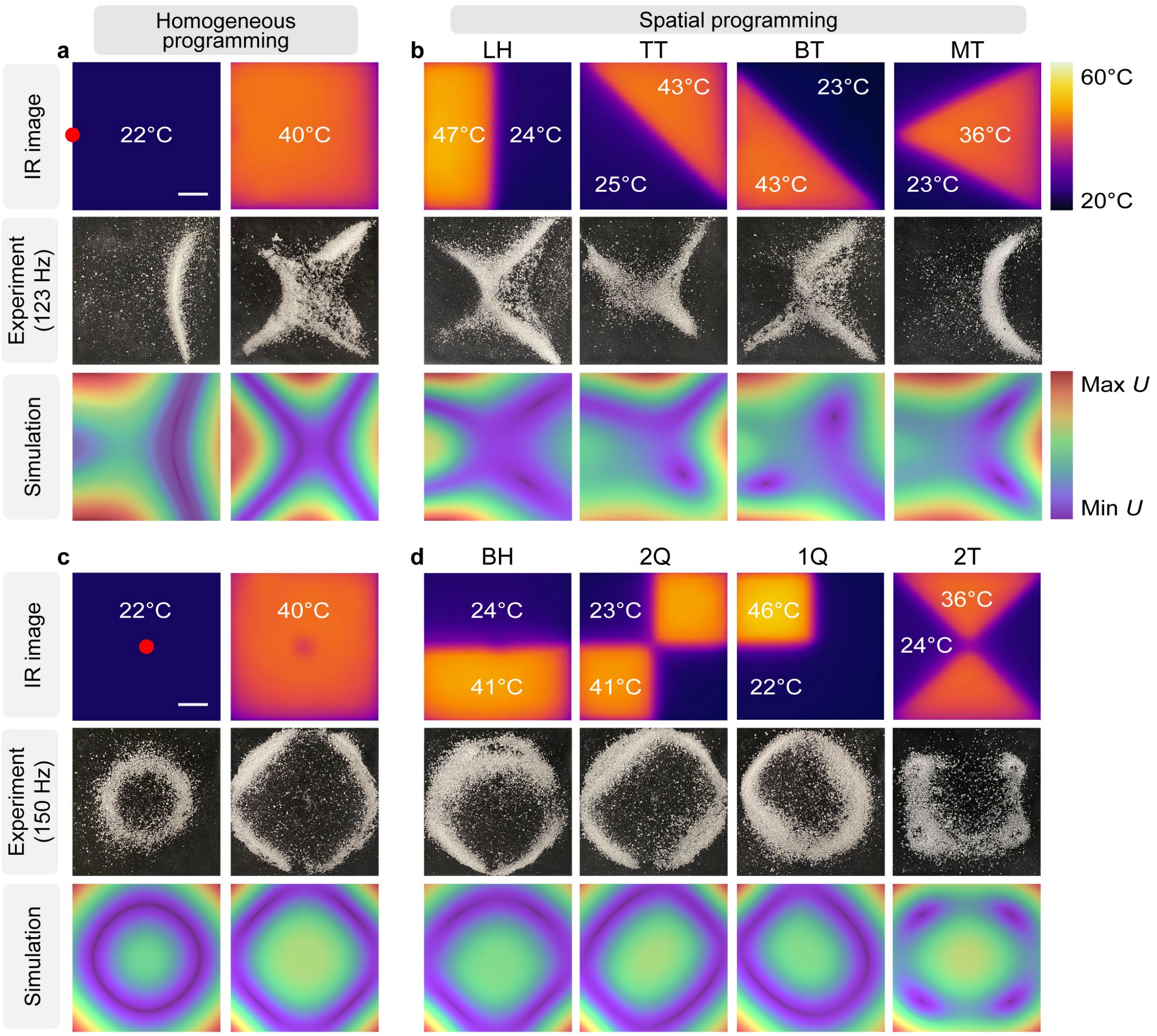

**Fig. 4: Spatial thermal programming of 2D vibrational modes.** **a** Homogeneous programming of a 100 mm × 100 mm × 1 mm C-SMP plate under cantilevered excitation (red dot). Top row: Infrared (IR) images of thermal programming. Middle row: experimental Chladni patterns. Bottom row: simulated displacement fields. **b** Spatial programming of a C-SMP plate under cantilevered excitation. Spatially selective actuation enables vibrational patterns not accessible through uniform heating. Patterns (left to right): LH, TT, BT, and MT. **c** Homogeneous programming of a C-SMP plate under center excitation (red dot). Top row: IR images of thermal programming. Middle row: experimental Chladni patterns. Bottom row: simulated displacement fields. **d** Spatial programming of a C-SMP plate under center excitation. Patterns (left to right): BH, 2Q, 1Q, and 2T. Scale bars: 20 mm.

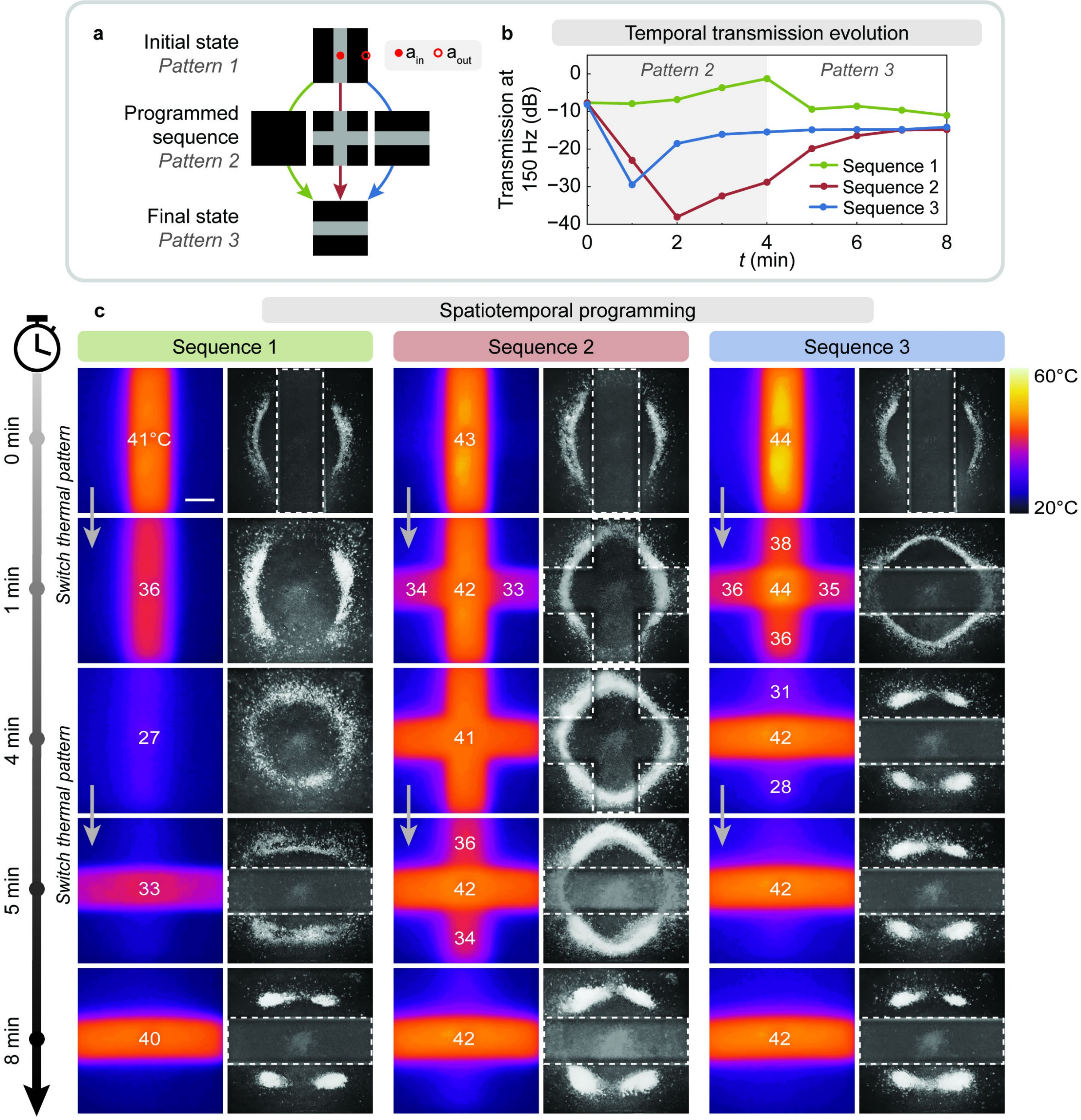

**Fig. 5: Spatiotemporal programming of 2D vibrational modes. a** Spatiotemporal programming of a 100 mm × 100 mm × 1 mm C-SMP plate under center excitation (red dot) at 150 Hz. The system transitions from a vertically illuminated state at 0 min to a horizontally illuminated state at 8 min through three programmed thermal sequences. From 0 to 4 minutes, intermediate illumination conditions consist of the light source being turned off (sequence 1, green), cross illumination (sequence 2, red), and horizontal illumination (sequence 3, blue). Black corresponds to regions with no light, and light gray corresponds to regions under irradiation. **b** Temporal evolution of the transmission at 150 Hz measured at the point $a_{out}$ (red circle) for the three thermal sequences shown in panel a. **c** Corresponding IR images and Chladni patterns illustrating the spatiotemporal evolution of the thermal and vibrational fields for each programmed sequence. Scale bar: 20 mm.